\documentclass[10pt,twocolumn]{article} 
\usepackage{simpleConference}
\usepackage{times}
\usepackage{graphicx}
\usepackage{amssymb}
\usepackage{url}
\usepackage{caption}
\usepackage{subcaption}	

\newtheorem{lemma}{Lemma}

\begin{document}

\title{Network Community Detection with A Successive Spectral Relaxation Method}

\author{Wenye Li \\
Shenzhen Research Institute of Big Data \\
The Chinese University of Hong Kong, Shenzhen \\
Dec. 1, 2017 \\
wyli@cuhk.edu.cn  \\
}

\maketitle
\thispagestyle{empty}

\begin{abstract}
With invaluable theoretical and practical benefits, the problem of partitioning networks for community structures has attracted significant research attention in scientific and engineering disciplines. In literature, Newman's \textit{modularity} measure is routinely applied to quantify the quality of a given partition, and thereby maximizing the measure provides a principled way of detecting communities in networks. Unfortunately, the exact optimization of the measure is computationally NP-complete and only applicable to very small networks. Approximation approaches have to be sought to scale to large networks. To address the computational issue, we proposed a new method to identify the partition decisions. Coupled with an iterative rounding strategy and a fast constrained power method, our work achieves tight and effective spectral relaxations. The proposed method was evaluated thoroughly on both real and synthetic networks. Compared with state-of-the-art approaches, the method obtained comparable, if not better, qualities. Meanwhile, it is highly suitable for parallel execution and reported a nearly linear improvement in running speed when increasing the number of computing nodes, which thereby provides a practical tool for partitioning very large networks.
\end{abstract}

\section{Introduction}
\label{sec:intro}

In graph theory, a network refers to a collection of vertices linked together via edges. It provides a powerful tool for modeling real-world entities as well as their complex interactions \cite{Newman:2010,EasleyKleinberg:2010}. Many problems can be investigated from this viewpoint. Specifically in scientific and engineering disciplines, transportation networks, communication networks, the World Wide Web, biological patterns, social connections, neural networks, metabolic networks, and pathological networks are all representative examples which can be modeled and studied from a network point of view \cite{JankowskiLorekJOW:2016}.

Networks in various domains exhibit diversity in forms. To understand the dynamics of these apparently different networks, the study of the invariant characteristics is necessary. Among the characteristics, community structures are widely believed to be common and important in real networks \cite{GirvanNewman:2002}. That is, the vertices fall naturally into groups with close intra-group relations but estranged inter-group relations. Identifying community structures helps distinguish the pairs of vertices that are more likely to be connected from those pairs that are less likely to be connected, which has both invaluable theoretical values and tremendous practical applications \cite{Li:INS:2015,Li:NEUCOM:2015} and becomes an imperative research topic.

To detect the underlying community structure for a given network, the ``modularity'' measure, developed by Girvan and Newman \cite{GirvanNewman:2002}, is routinely applied. The measure is expressed as the differences between the real fraction of edges connecting vertices within each community and the expected fraction when the edges were assumed to be uniformly distributed. It has been shown by extensive studies on both empirical and simulated networks that larger modularity values usually leads to better vertex partitions. Thereby maximizing the modularity measure provides a mathematically well-posed approach in revealing the underlying community structures in networks \cite{MuchaRMPO:2010,FacchettiIA:2011,SzellaLT:2010,Li:INS:2013}.

Unfortunately, optimizing the modularity measure is mathematically difficult. It is known that the exactly maximizing the measure is an NP-complete problem over all graphs of a given size, and is only feasible for networks with up to a few hundred vertices \cite{Brandes06}. For larger networks, approximate solutions have to be sought to ensure the scalability, often at the price of losing accuracy. Despite the limited success that has been achieved by the state-of-the-art methods, it is highly desirable to work out a method that is applicable to large networks with high accuracy.

Our work developed a novel spectral relaxation based method to maximize the modularity measure. Coupled with an iterative rounding strategy and a simple constrained power method, it provides a fast solution for detecting communities in large networks with high qualities. Another key benefit of the method is that it mainly involves basic matrix-vector operations, which can be easily performed in parallel with high efficiency. The method was implemented and tested on a parallel computing cluster with $128$ CPU cores. A nearly linear improvement in running speed was observed when increasing the number of computing nodes, which strongly verified the potential of the method in partitioning very large networks.

The rest of this paper is structured as follows. Section 2 introduces the background knowledge including the modularity maximization model and its related work. Section 3 illustrates our successive spectral relaxation based approach in detail. Section 4 reports the empirical evaluation results, followed by the conclusion in Section 5.

\section{Background}
\label{sec:background}

\subsection{Graph partition and community detection}

Graph partition and community detection are two related problems yet with significant difference. Graph partition often arises in computer science, mathematics and physics. The problem is well-defined and has been studied since the 1960s \cite{KernighanLin:1970}. It usually refers to the task of splitting the vertices of a network into groups of fixed numbers or of given sizes with the objective of minimizing the number of edge connections between groups.

Comparatively community detection is a much newer problem and has been studied mainly in the recent decade, but has appeared in much wider areas of natural sciences and social sciences including physics, chemistry, biology, social networks, and so on. In community detection, the number of groups and the size of each group are not specified in advance, but are determined by the network itself. The objective is to find a ``natural fault line'' along which a given network divides into partitions \cite{Newman:2010}.

A number of community detection models and computational methods have been developed in literature \cite{Newman:2010}. These models try to address different aspects of networks and lead to different computer algorithms. In this paper, out work focuses on the modularity model and proposes an effective algorithm that runs efficiently on parallel computing platforms.

\subsection{Modularity maximization on undirected networks}
\label{sec:modularity:undirected}

For a candidate division of a network, the \emph{modularity} measure is routinely applied to quantify the quality of the partition. Good divisions, with high modularity values, have dense intra-community connections (edges between vertices in the same group) but sparse inter-community connections (edges between vertices in different groups). The modularity measure expresses the concentration of edges inside each group compared with a uniform distribution of edges between each pair of vertices regardless of group partitions.

Let us start the discussion from a simplified case. Assume $G=\left( V,E \right)$ is an undirected network, with a set of vertices $V=\left\{ v_1,v_2,\cdots,v_n \right\}$ and a set of undirected edges $E$. Let $a_{ij}=1 $ if there is an edge connecting $v_i$ and $v_j$, and $a_{ij}=0$ otherwise. For each vertex $v_i$, denote by $d_{i}=\sum_{j=1}^{n}a_{ij}$ its degree. Also denote by $m=\frac{1}{2}\sum_{i=1}^{n}d_{i}$ the total number of edges in the network.

Given a candidate assignment of network vertices into groups, the modularity model assumes that the degree associated with each vertex holds preserved. With the uniform random selection principle, we know that the expected number of connections between any two vertices $v_i$ and $v_j$ is $\frac{d_{i}d_{j}}{2m}$. Therefore the real observation minus the expectation is given by $a_{ij}-\frac{d_{i}d_{j}}{2m}$. Sum over all pairs of vertices in the same group. The modularity measure, denoted by $Q$, is defined by%
\begin{equation}
Q=\frac{1}{2m}\sum_{i,j=1}^{n}\left[ a_{ij}-\frac{d_{i}d_{j}}{2m}\right] \delta_{ij}
\label{equ:modularity:undirected}
\end{equation}%
with $\delta_{ij}=1$ if the vertices $v_i$ and $v_j$ are assigned in the same group and $\delta_{ij}=0$ otherwise.

The modularity value has a range $\left[-\frac{1}{2},1\right)$. It is positive when the observed number of intra-group edges is greater than the expectation on the basis of chance. It has been verified through numerous real and simulated studies that larger modularity values are correlated with better community structures in networks. Therefore, optimizing the modularity measure provides a practical and principled way of partitioning networks. Through searching for the partition that has the largest modularity value, one can detect community structures in networks precisely.

\subsection{Modularity maximization on directed networks}
\label{sec:modularity:directed}

With trivial modification, the modularity model on undirected networks can be applied in the case of directed networks as well \cite{LeichtNewman:2008}, with which the vertices typically have different in-degrees and out-degrees. Consider a directed network with $n$ vertices $\left\{ v_1,v_2,\cdots,v_n \right\}$. It has an edge from vertex $v_{j}$ to vertex $v_{i}$ with probability $\frac{d^{in}_{i}}{d^{out}_{j}}$, where $d^{in}_{i}$ is the in-degree of $d_i$ and $d^{out}_{j}$ is the out-degree of $d_j$. Let $a_{ij}=1$ if there is a directed edge from $v_j$ to $v_i$ and $a_{ij}=0$ otherwise. Similarly denote by $m$ the number of directed edges in the network. Then the modularity measure on the directed network is defined by:
\begin{equation}
Q=\frac{1}{m}\sum_{i,j=1}^{n}\left[ a_{ij}-\frac{d^{in}_{i}d^{out}_{j}}{m}\right] \delta_{ij}.
\label{equ:modularity:directed}
\end{equation}%
Note that, different from the modularity measure on undirected networks, there is no factor of $2$ in the denominator of the model.

The modularity models on both undirected and directed networks can be extended to the case of weighted networks, which can be done trivially by replacing $a_{ij}$ to be the edge weight if there is an edge between $v_i$ and $v_j$. Here we omit the detailed discussion.

\subsection{Modularity maximization methods}
\label{sec:modularity:methods}

Partitioning networks by maximizing the modularity measure exactly is a known NP-complete problem \cite{Brandes06}. The required computation grows exponentially with the increasing size of the network. Despite the challenge of NP-completeness, a number of exact methods for exhaustive optimization were developed, which achieved limited success on networks with up to a few hundred vertices on conventional computing platforms, such as the integer programming approach and the column generation method \cite{AgarwalKempe:2008,AloiseCCHPL:2010}.

To ensure the tractability on large networks, approximate algorithms have to be sought. Through relaxation, Agarwal \& Kempe designed a linear program method \cite{AgarwalKempe:2008}. On small networks, the method reported very accurate results. Unfortunately, the method is still computationally demanding and does not scale to large networks.

The simulated annealing method was investigated in the problem \cite{GuimeraAmaral:2005}. Simulated annealing treats the quantity of interest as an energy and simulates the cooling process of solids until the system reaches the state with the lowest energy. The method had excellent empirical performance and reported the best known results on many real networks. Unfortunately, although the method partially lessens the computational requirement, the burden is still prohibitive for very large networks.

Greedy heuristics were investigated on large networks. A straightforward way is to start with each vertex in a group of its own. The method then successively combines a pair of groups into one group. At each step it chooses the two groups with which the combination gives the largest modularity value increase, or the smallest decrease if no choice gives an increase. Eventually all vertices are merged into a single group. Then we go back over all the intermediate steps, select the one state with the highest modularity value and obtain the partition result \cite{CalusetNM:2004,WakitaTsurumi:2007}.

A related heuristic is based on edge \textit{betweenness} \cite{GirvanNewman:2002}. For an edge, its betweenness is given by the number of shortest paths of all pairs of vertices that pass through the edge. The heuristic recursively seeks and removes one by one the edges with the highest betweenness, until the network breaks up into single vertices. Therefore the procedure generates a dendrogram with hierarchical divisions from a single group to all isolated vertices, with which the intermediate division possessing the highest modularity value will be chosen. Overall these greedy methods run fast, and give moderately good divisions of networks. But in practice the two simple heuristics have been superseded by alternatives that often find higher modularity values \cite{Newman:2010}.

More complicated search heuristics were specially designed for the modularity maximization problem. Recently, Noack \& Rotta developed a multi-level search method \cite{RottaNoack:2011}, which involves coarsening- and refinement-based heuristics. Another method, the Louvain method \cite{BlondelGLL:2008}, uses a two-phase iterative search strategy by firstly looking for small communities through optimizing modularity locally and then aggregating nodes in the same community to construct a new network. These two search heuristics have reported very accurate results on many benchmarked networks and are regarded as state-of-the-art solutions for partitioning large networks \cite{AynaudBGL:2013}.

\section{A Successive Spectral Relaxation Method}
\label{sec:sar}

\subsection{Conventional spectral relaxation}
\label{sec:sar:conventional}

The idea of spectral relaxation can be applied in community detection problems \cite{Newman:2006,WhiteSmyth:2005}. To illustrate the method, let us start from a special case of dividing an undirected network into just two groups. Use $s_{i}=\pm 1$ to denote the group membership of vertex $v_i$. Then we have $\sum_{i}s^{2}_{i}=n$ and $\delta_{ij}=\frac{1}{2}\left(s_{i}s_{j}+1\right)$. Then
\begin{equation}
Q=\frac{1}{4m}\sum_{i,j=1}^{n}\left[a_{ij}-\frac{d_{i}d_{j}}{2m}\right]\left(s_{i}s_{j}+1\right)=\frac{1}{4m}s^{T}Bs
\label{equ:modularity:binary:undirected1}
\end{equation}
where $B$ is an $n\times n$ \textit{modularity} matrix with elements $b_{ij}=a_{ij}-\frac{d_{i}d_{j}}{2m}$. The sums of elements in each row and in each column of $B$ are all zero, which implies that the modularity value of an un-divided network is always zero.

Label all eigenvalues of the modularity matrix in a non-increasing order $\lambda_{1}\geq \lambda_{2}\geq \cdots \geq \lambda_{n}$. Assume $u_{i}$ is the unit eigenvector associated with eigenvalue $\lambda_{i}$. Then the vector $s=\sum_{i}a_{i}u_{i}$, where $a_{i}=u_{i}^{T}s$. And we have
\begin{equation}
Q=\frac{1}{4m}\sum_{i=1}^{n}a_{i}u_{i}^{T}B\sum_{j=1}^{n}a_{j}u_{j} =\frac{1}{4m}\sum_{i=1}^{n}\left(u_{i}^{T}s\right)^{2}\lambda_{i}
\label{equ:modularity:binary:undirected2}
\end{equation}

To maximize the value of $Q$, it is obvious that the vector $s$ needs to be chosen in a way such that as much weight as possible is concentrated involving the largest eigenvalue $\lambda_{1}$. Correspondingly the best choice of $s$ should be proportional to the first eigenvector $u_1$. Unfortunately with the constraint that each element of $s$ only takes the value of $1$ or $-1$, such a proportion is generally infeasible, which makes the optimization process a hard problem.

A simple rounding strategy is often applied and found effective in practice, with which the vertices are divided into two groups based on the signs of the elements in the eigenvector $u_1$. That is: $s_{i}=+1$ if $u_{1i}>0$ and $s_{i}=-1$ otherwise, where $u_{1i}$ is the $i$-th element of $u_1$.

Now the network partition problem is simplified to the problem of estimating the eigenvector $u_{1}$ of the modularity matrix $B$. The eigenvector can be calculated by the power iteration method efficiently. Starting with a random vector $v^0$, the power iteration method updates the vector through matrix-vector multiplication and normalization:
\begin{equation}
v^{i+1}=\frac{Bv^{i}}{\left\| Bv^{i} \right\|},
\end{equation}
with $\left\| \cdot \right\|$ denoting the $\ell_2$-norm of a vector. After a number of iterations, the process gradually approaches the dominant eigenvector $v$, which is the eigenvector associated with the dominant eigenvalue $\lambda$ that has the largest magnitude. 

If the dominant eigenvalue $\lambda>0$, it is the first eigenvalue $\lambda_1$ and the dominant eigenvector $v$ is just the desired eigenvector $u_1$. The vertices are then divided into two communities based on the signs of the elements in $u_1$.

If the dominant eigenvalue $\lambda<0$, however, it is $\lambda_n$ and the dominant eigenvector $v$ is $u_n$ instead of the desired $u_1$. In this case, we can shift the matrix $B$ to: $B^{\prime}=B+\left| \lambda \right|I$, where $I$ is the identity matrix of the same size as $B$. $B^{\prime}$ has the eigenvalues $\lambda_{1}+\left| \lambda \right|\ge \lambda_{2}+\left| \lambda \right|\ge\cdots \ge\lambda_{n}+\left| \lambda \right|$ and the same eigenvectors $u_1,u_2,\cdots,u_n$ as $B$. But applying the power iteration method on $B^{\prime}$ returns the desired dominant eigenvector $u_1$.

When a directed network needs to be divided into two communities, again we define $s_{i}=+1$ if vertex $v_{i}$ is to be assigned to one community and $s_{i}=-1$ otherwise, which similarly leads to the maximization of
\begin{equation}
Q=\frac{1}{2m}\sum_{i,j=1}^{n}s_{i}b_{ij}s_{j}=\frac{1}{2m}s^{T}Bs
\label{equ:modularity:binary:directed}
\end{equation}
with respect to $s\in \left\{-1,+1\right\}^n$, where the matrix $B=\left(b_{ij}\right)_{i,j=1}^{n}$ and $b_{ij}=a_{ij}-\frac{d^{in}_{i}d^{out}_{j}}{m}$.

The modularity matrix $B$ in the case of directed networks is, in general, not symmetric. To restore the symmetry, we maximize
\[
Q=\frac{1}{4m}s^{T}\left(B+B^{T}\right)s
\]
instead. Similarly the spectral relaxation method can be applied based on the first eigenvector of the matrix $B+B^{T}$.

For network partition into more than two groups, this two-way division scheme is performed recursively on each group \cite{Newman:2010,LiSchuurmans:2011}. The division process repeats until there is no increase in $Q$'s value, which happens when the modularity matrix has no positive eigenvalues.

\subsection{Successive relaxation and the constrained power method}
\label{sec:sar:cpm}

The conventional spectral relaxation method discussed in Section \ref{sec:sar:conventional} divides network vertices into two partitions according to the signs of the elements in the first eigenvector of the modularity matrix, while completely ignoring their magnitudes. However, the magnitudes contain important information. It is evident from Equ. (\ref{equ:modularity:binary:undirected2}) that a large magnitude would contribute significantly to the modularity value and therefore give us strong confidence in deciding the group membership of the corresponding vertex. Contrarily, a small magnitude makes it difficult to set the membership of the vertex due to its trivial influence on the modularity value.

Considering the important information the magnitudes have, it is intuitively desirable and technically feasible to take the magnitudes into consideration and design a \textit{successive relaxation} method for network partition. Initially, the successive relaxation method is the same as the conventional relaxation approach and applies the power iteration method to compute the first eigenvector of the modularity matrix. The difference is, rather than making the division decision in a single batch, we only set the group membership of the vertices with large magnitudes. The decision of the remaining vertices with small magnitudes are postponed. In the forthcoming iterations, a residual problem is generated. The structure of the new problem is roughly the same as the first one but with fewer un-partitioned vertices and we can deal with it in a similar way. The process is repeated until no vertices are left un-partitioned.

Mathematically, in the first iteration the spectral relaxation method solves $\max s^{T}Bs$, the same problem as in Section \ref{sec:sar:conventional}. Again we apply the classical power method to obtain the first eigenvector $u_1$ of the modularity matrix $B$. Then, instead of deploying the conventional rounding strategy, partition decisions are only made on those elements with sufficiently large magnitudes, i.e.,
\begin{equation}
s_{i}=\left\{
\begin{array}{l}
+1 \\
-1 \\
unknown%
\end{array}%
\right.
\begin{array}{l}
if \enspace u_{1i}\ge\sigma  \\
if \enspace u_{1i}\le-\sigma  \\
otherwise%
\end{array}%
\label{equ:ssr:threshold}
\end{equation}
where $\sigma$ is a positive threshold value, often setting as one.

Denote by $s_+$ the rounded elements whose values have been held fixed in the first iteration, and $s_-$ the remaining elements awaiting to be set. Re-organize
$s=\left(
\begin{array}{c}
s_{+} \\
s_{-}%
\end{array}%
\right)$.
The new optimization objective becomes%
\begin{equation}
\left(
\begin{array}{c}
s_{+} \\
s_{-}%
\end{array}%
\right) ^{T}\left(
\begin{array}{cc}
B_{++} & B_{+-} \\
B_{-+} & B_{--}%
\end{array}%
\right) \left(
\begin{array}{c}
s_{+} \\
s_{-}%
\end{array}%
\right)
\end{equation}%
where $B_{++},B_{+-},B_{-+}$ and $B_{--}$ are four submatrices of $B$. Note that the value of $s_{+}^{T}B_{++}s_{+}$ holds constant and thus can be ignored. The objective becomes equivalently the maximization of%
\begin{equation}
L=s_{-}^{T}B_{--}s_{-}+2s_{-}^{T}B_{-+}s_{+}
\label{equ:cpm:objective}
\end{equation}%
with respect to $s_{-}$, subject to the length constraint: $\left\Vert s_{-}\right\Vert = \sqrt{k}$ where $k$ denotes the number of elements in $s_{-}$.

To solve the new problem, we designed a \textit{constrained power method} that has a simple update rule:%
\begin{equation}
s_{-}^{i+1}=\frac{B_{--}s_{-}^{i}+B_{-+}s_{+}}{\left\Vert
B_{--}s_{-}^{i}+B_{-+}s_{+}\right\Vert } \times \sqrt{k}.
\label{equ:cpm:update}
\end{equation}

The update rule can be intuitively explained from a viewpoint of gradients. Maximizing $L$ requires updating $s_{-}$ along the gradient direction and re-normalizing the vector to satisfy the norm constraint. The update is guaranteed to converge, which happens when the gradient direction $\nabla L$ is parallel to (propositional to) the current estimate of $s_{-}$: $\nabla L \varpropto s_{-}$. By taking the derivative of $L$ with respect to $s_{-}$, we also know $\nabla L \varpropto B_{--}s_{-}+B_{-+}s_{+}$. Therefore, it holds that $s_{-} \varpropto B_{--}s_{-}+B_{-+}s_{+}$. Considering that $s_{-}$ has a length of $\sqrt{k}$ and $\frac{B_{--}s_{-}^{i}+B_{-+}s_{+}}{\left\Vert B_{--}s_{-}^{i}+B_{-+}s_{+}\right\Vert }$ has a unit length, then $s_{-}=\pm \frac{B_{--}s_{-}+B_{-+}s_{+}}{\left\Vert B_{--}s_{-}+B_{-+}s_{+}\right\Vert } \times \sqrt{k}$. Taking the positive one, we have the update rule in Equ. (\ref{equ:cpm:update}).

As in the first iteration, given the relaxed solution of $s_{-}$, a similar partial rounding procedure is adopted. Only those elements with sufficiently large magnitudes are rounded and fixed. In this way, the iterative rounding procedure and the constrained power method are performed successively to determine the group membership of the vertices. The process tops when $s_{-}$ becomes empty when all vertices have been allocated into two groups.

\subsection{Complexity}
\label{sec:sar:complexity}

For a given network with $n$ vertices and $m\le kn$ edges where $k$ is a constant, a known result, based on the work of \cite{Newman:2010}, is that the classical power method needs $O\left(n \right)$ matrix-vector multiplications to calculate the leading eigenvector of the modularity matrix and each multiplication needs $O\left(n \right)$ floating point operations by taking the sparsity of the network into consideration. So in total, the spectral relaxation method needs $O\left(n^2\right)$ operations to partition a network into two groups based on the modularity model.

The complexity of the successive spectral relaxation method can be analyzed similarly with a small modification. Instead of using a threshold value $\sigma$ as in Equ. (\ref{equ:ssr:threshold}) to determine the borderline of rounding, we assume an $\epsilon \left(0<\epsilon <1\right)$ faction of unrounded vertices' group membership get decided in each iteration. Similarly to the power method, in the first iteration the constrained power method has a complexity of $O\left(n^2\right)$ for a sparse network with $n$ variables. In the subsequent iteration the residual problem has $n\left(1-\epsilon \right)$ unrounded vertices, and the constrained power method would therefore require $O\left(n^2\left(1-\epsilon\right)^2\right)$ floating operations to converge. Repeating the argument, the complexity of the successive spectral relaxation method is given by:
\[
n^2+n^2\left(1-\epsilon\right)^2+n^2\left(1-\epsilon\right)^4+\cdots = \frac{1}{2\epsilon-\epsilon^2}n^2,
\]
which lies between $\frac{1}{2\epsilon}n^2$ and $\frac{1}{\epsilon}n^2$. So we have the following result:
\begin{lemma}
To bipartition a network with $n$ vertices and $m\le kn$ edges where $k$ is a constant, the successive spectral relaxation method has a complexity of $O\left(\frac{1}{\epsilon}n^2 \right)$ where $\epsilon$ is the fraction of variables to round in each iteration.
\end{lemma}

\subsection{Relationship with the projected power method}

\begin{figure}[!t]
\centering
\includegraphics[width=3.2in]{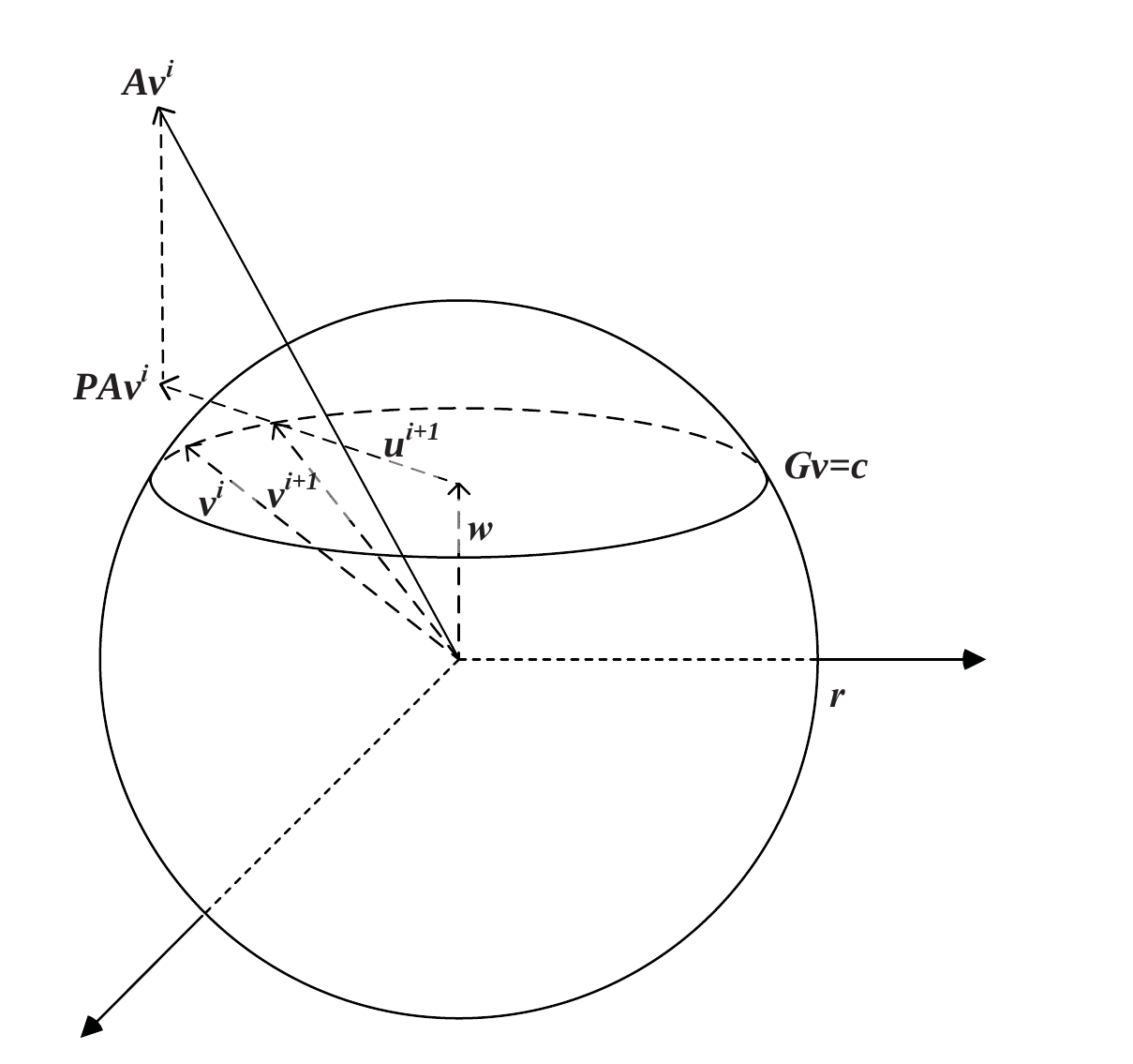}
\caption{A graphical illustration of the projected power method.}
\label{fig:ppm}
\end{figure}

The proposed constrained power method can be derived rigorously from the projected power method \cite{XuLS:2009}, which investigates a generic optimization problem,
\begin{equation}
\max_v v^{T}Av \quad \mbox{ subject to } \quad
\left\Vert v \right\Vert=r, Gv=c.
\label{equ:ppm:problem}
\end{equation}
where $A$ is a positive definite matrix, $r$ is a positive value, and $Gv=c$ denotes the linear constraints exerted on $v$.

As shown in Fig. \ref{fig:ppm}, all feasible solutions of $v$ to the maximization problem are vectors starting from the origin and ending on the surface of $\left\Vert v \right\Vert=r$. Let $w$ be the vector from the origin to its projection point on the hyperplane $Gv=c$. It can be easily seen that every feasible solution of $v$ can be written as $v=u+w$ where vector $u$ lies on the hyperplane $Gu=0$ and $\left\Vert u \right\Vert=\sqrt{r^2-w^{T}w}$. The projection of vector $v$ onto the hyperplane $Gv=c$ is given by $Pv$, where $P=I-G^{T}\left(GG^{T} \right)^{-1}G$ is the projection matrix and $I$ denotes the identity matrix of appropriate size.

In each iteration, given the current $v^{i}$, the projected power method stretches the vector by multiplying with $A$, projects the stretched vector $Av^{i}$ onto $Gv=c$, and re-normalizes the projection to $u^{i+1}$. Finally we obtain $v^{i+1}$ by summing up $u^{i+1}$ and $w$. That is, the projected power method has an update rule of:
\begin{equation}
v^{i+1}=\frac{PAv^i}{\left\Vert PAv^i \right\Vert}\times \sqrt{r^2-w^{T}w} + w.
\label{equ:ppm:update}
\end{equation}
It can be proved that during each step, the estimate of $v$ gets nearer and nearer to the maximum stretching direction of $A$ while staying feasible. The convergence is theoretically guaranteed, and the convergence speed is usually very fast in practice \cite{XuLS:2009}.

The update rule of the constrained power method in Equ. (\ref{equ:cpm:update}) can be rigorously derived from the update rule of the projected power method in Equ. (\ref{equ:ppm:update}). Without loss of generality, we just assume $B_{--}$ is a positive definite matrix\footnote{If $B_{--}$ is not positive definite, its diagonal elements can be shifted by a positive value to provide the positive definiteness, as shown in Section \ref{sec:sar:conventional}.}. We then re-write the optimization objective in Equ. (\ref{equ:cpm:objective}) as
\begin{equation}
L=v^{T}Av-z
\label{equ:maxobj2}
\end{equation}
where $A=\left[
\begin{array}{cc}
B_{--} & B_{-+}s_{+} \\
\left( B_{-+}s_{+}\right) ^{T} & z%
\end{array}%
\right]
$ and $v=\left[
\begin{array}{c}
s_- \\
1%
\end{array}%
\right]$.
With a sufficiently large value of $z$, the positive definiteness of matrix $A$ can be ensured. When $z$ is given, the objective becomes equivalently the maximization of $v^{T}Av$ satisfying: $\left\Vert v \right\Vert=\sqrt{k+1}$ and $v_{k+1}=1$.

By exploring the structure of the problem in Equ. (\ref{equ:maxobj2}, we are able to get a simple solution by applying the update rule of the projected power method in Equ. (\ref{equ:ppm:update}). Decompose a feasible solution $v$ into $v=u+w$ where $w=\left[0,\cdots,0,1 \right]^{T}$ and $u$ is a vector satisfying $u_{k+1}=0$ and $\left\Vert u \right\Vert=\sqrt{k}$. Given the estimate in the $i$-th iteration $v^i=\left[
\begin{array}{c}
s_-^i \\
1%
\end{array}%
\right]$, we stretch it to $Av^{i}=\left[
\begin{array}{c}
B_{--}s_{-}^{i}+B_{-+}s_{+} \\
\left( B_{-+}s_{+}\right) ^{T}s_{-}^{i}+d %
\end{array}%
\right]$. Project the vector onto the hyperplane $u_{k+1}=0$, and it becomes $\left[
\begin{array}{c}
B_{--}s_{-}^{i}+B_{-+}s_{+} \\
0 %
\end{array}%
\right]$. Re-normalize the result and we have a new estimate $v^{i+1}=
\left[
\begin{array}{c}
s_{-}^{i+1}\\
1
\end{array}
\right]=
\left[
\begin{array}{c}
\frac{B_{--}s_{-}^{i}+B_{-+}s_{+}}{\left\Vert B_{--}s_{-}^{i}+B_{-+}s_{+}\right\Vert} \times \sqrt{k}\\
1 %
\end{array}%
\right]$ by summing up $u^{i+1}$ and $w$, which exactly gives the update rule of the constrained power method in Equ. (\ref{equ:cpm:update}).

\subsection{Parallelizability}
\label{sec:sar:parallel}

Parallel computing refers to the type of computation with which the calculations are carried out simultaneously on multiple computing nodes \cite{Quinn:1994}. It has been employed for decades, mainly in high-performance computing, and has helped solve many difficult problems that cannot be tackled by conventional serial computing models. Nowadays, parallel computing is becoming more and more important in handling large-scale data processing applications.

A key concern to the success of parallel computing is to divide the execution of an algorithm into parallel portions that can be distributed and solved independently. Practically, the algorithms are very different in the level of parallelizability, varying from easily parallelizable to totally unparallelizable at all. Another concern lies in the communication and synchronization costs for different computing nodes, which also affect the parallelizability of an algorithm significantly. 

The constrained power method proposed in this paper can be parallelized easily and effectively. The method runs iteratively, and in each iteration it mainly involves matrix-vector multiplication and addition operations. The matrix operands are easily split into smaller blocks so that the operations on each block can be executed simultaneously on different computing nodes. The final result is obtained by merging results from all blocks with small communication and synchronization costs that can be neglected. As a result, the proposed method has high efficiency in parallel execution. Empirically in our evaluation, a nearly linear improvement in running speed was observed when increasing the number of computing nodes.

\section{Evaluation}
\label{sec:evaluation}

We evaluated the proposed method thoroughly on both real and synthetic networks, with three objectives: to evaluate the partition quality (i.e. the modularity values) with real networks, to evaluate the method's sensitivity towards the change of structures with synthetic networks, and to evaluate the method's running speed and parallel execution efficiency with a very large networks.

\subsection{Modularity values on real networks}
\label{sec:evaluation:modularity}

Eighteen networks were used to evaluate and compare the empirical performance of different partition methods in modularity values. These networks, listed in Table \ref{tab:networks}, are from two collections publicly available in the Internet: from Mark Newman's website\footnote{http://www-personal.umich.edu/$\sim$mejn/netdata/} and from Stanford large network dataset collection\footnote{https://snap.stanford.edu/data/} \cite{LeskvecKrevl:2014}. The two collections include directed/undirected and weighted/unweighted networks and cover a wide range of real applications including social networks, co-purchase networks, email networks, cooperation networks, citation networks, product networks, etc. The sizes of the networks vary significantly from less than one hundred vertices and edges, to over three million vertices and sixteen million edges. In literature, these networks have been popularly used as benchmarks in evaluating community detection algorithms.

\begin{table*}[!t]
\caption{Benchmark networks from Mark Newman's personal website and Stanford large network dataset collection.}
\label{tab:networks}
\begin{center}
\begin{small}
\begin{tabular}{lccl}
\hline
Networks & $\#(vertices)$ & $\#(edges)$ & Description \\ \hline
karate    & $34$  & $78$ & Friendship relations of members in a karate club \\
dolphins  & $62$  & $159$ & Frequent associations of dolphins \\
lesmis    & $77$  & $254$ & Character interactions from \textit{Les Mis\'erables} \\
polbooks  & $105$ & $441$ & Co-purchase of politics books from \textit{Amazon.com} \\
adjnoun   & $112$ & $425$ & Adjacency of adjectives and nouns in \textit{David Copperfield} \\
football  & $115$ & $613$ & American college football games network (2000) \\
jazz    & $198$ & $2,742$ & Jazz musicians network \\
email     & $1,133$ & $5451$ & An email communication network \\
ca-GrQc   & $5,242$ & $28,980$ & Collaboration net of arxiv general relativity \\
ca-HepTh  & $9,877$ & $51,971$ & Collaboration net of arxiv high energy physics theory \\
ca-HepPh  & $12,008$ & $237,010$ & Collaboration network of arxiv high energy physics \\
ca-AstroPh  & $18,772$ & $396,160$ & Collaboration net of arxiv astro physics \\
ca-CondMat  & $23,133$ & $186,936$ & Collaboration net of arxiv condensed matter \\
cit-HepTh   & $27,770$ & $352,807$ & Paper citation net of arxiv high energy physics theory \\
cit-HepPh   & $34,546$ & $421,578$ & Paper citation net of arxiv high energy physics \\
com-DBLP  & $317,080$ & $1,049,886$ & DBLP collaboration network \\
com-Amazon  & $334,863$ & $925,872$ & Amazon product network \\
cit-Patents & $3,774,768$ & $16,518,948$ & US patent citation network (1975-1999) \\
\hline
\end{tabular}
\end{small}
\end{center}
\vskip -0.1in
\end{table*}

\begin{table}[!t]
\caption{Comparison of modularity values obtained by different methods. (For computational concerns, SA used an annealing parameter value of $0.99$ on networks with less than $5,000$ vertices, and a value of $0.90$ on networks with more than $5,000$ vertices.)}
\label{tab:qvalues}
\begin{center}
\begin{small}
\begin{tabular}{lcccccc}
\hline
Networks & CG & LP & SA & MLS & LOU & SSR \\
\hline
karate    & $.420$ & $.420$ & $.420$ & $.420$ & $.420$ & $.420$ \\
dolphins  & $.529$ & $.529$ & $.527$ & $.528$ & $.527$ & $.527$ \\
lesmis    & $.560$ & $.560$ & $.556$ & $.557$ & $.560$ & $.560$ \\
polbooks  & $.527$ & $.527$ & $.527$ & $.527$ & $.527$ & $.527$ \\
adjnoun   & $.308$ & $.308$ & $.308$ & $.308$ & $.308$ & $.308$ \\
football  & $.605$ & $.605$ & $.604$ & $.605$ & $.605$ & $.605$ \\
jazz    & $.445$ & $.445$ & $.445$ & $.445$ & $.445$ & $.445$ \\
email     & $-$ & $-$ & $.575$ & $.575$ & $.576$ & $.576$ \\
ca-GrQc     & $-$ & $-$ & $.853$ & $.861$ & $.863$ & $.863$ \\
ca-HepTh    & $-$ & $-$ & $.765$ & $.770$ & $.770$ & $.770$ \\
ca-HepPh    & $-$ & $-$ & $.640$ & $.657$ & $.658$ & $.663$ \\
ca-AstroPh  & $-$ & $-$ & $.609$ & $.627$ & $.622$ & $.630$ \\
ca-CondMat  & $-$ & $-$ & $.712$ & $.729$ & $.730$ & $.734$ \\
cit-HepTh   & $-$ & $-$ & $.630$ & $.656$ & $.659$ & $.658$ \\
cit-HepPh   & $-$ & $-$ & $.709$ & $.725$ & $.726$ & $.729$ \\
com-DBLP    & $-$ & $-$ & $-$ & $-$ & $.822$ & $.819$ \\
com-Amazon  & $-$ & $-$ & $-$ & $-$ & $.925$ & $.927$ \\
cit-Patents & $-$ & $-$ & $-$ & $-$ & $.810$ & $.813$ \\
\hline
\end{tabular}
\end{small}
\end{center}
\vskip -0.1in
\end{table}

The performance of the proposed successive spectral relaxation method (denoted by SSR) was compared with several state-of-the-art algorithms, including the linear programming method (LP) of Agarwal and Kempe \cite{AgarwalKempe:2008}, the simulated annealing method (SA) of Guimer{\`a} and Amaral \cite{GuimeraAmaral:2005}, the multi-level search method (MLS) of Noack \& Rotta \cite{NoackRotta:2008}, and the Louvain method (LOU) of Blondel et al. \cite{BlondelGLL:2008}. Besides, the optimal results obtained from the column generation method (CG) Aloise et al. \cite{AloiseCCHPL:2010}, which finds the optimal solution but only works on networks with up to a few hundred vertices, is also included as a reference when available.

Table \ref{tab:qvalues} compares the modularity values obtained by different methods. On small networks with less than $1,000$ vertices with which the optimal modularity values are known by the CG method, all methods reported highly effective results that were equal to or at least very near to the optimal values. 

On networks with more than $1,000$ vertices, there are no known optimal modularity values due to the prohibitive computation required by the exact methods. Besides, some approximation methods may also require huge amounts of computation. For example on most of the networks, the LP method couldn't finish the execution within $24$ hours on our platform and the corresponding results were therefore left blank in Table \ref{tab:qvalues}.

Among all results that are available on networks with more than $1,000$ vertices, the MLS, LOU and SSR methods reported very similar modularity values. Comparably the results of the SA method seemed to be inferior. One possible reason is that, to lessen the computation, the SA method used an annealing parameter value of $0.90$ when partitioning networks with more than $5,000$ vertices, rather than using the value of $0.99$ when partitioning smaller networks.

\subsection{Sensitivity on Synthetic Networks}
\label{sec:evaluation:sensitivity}

\begin{figure}[!t]
  \centering
  \begin{subfigure}[t]{2.5in}
    \centering
    \includegraphics[width=2.5in]{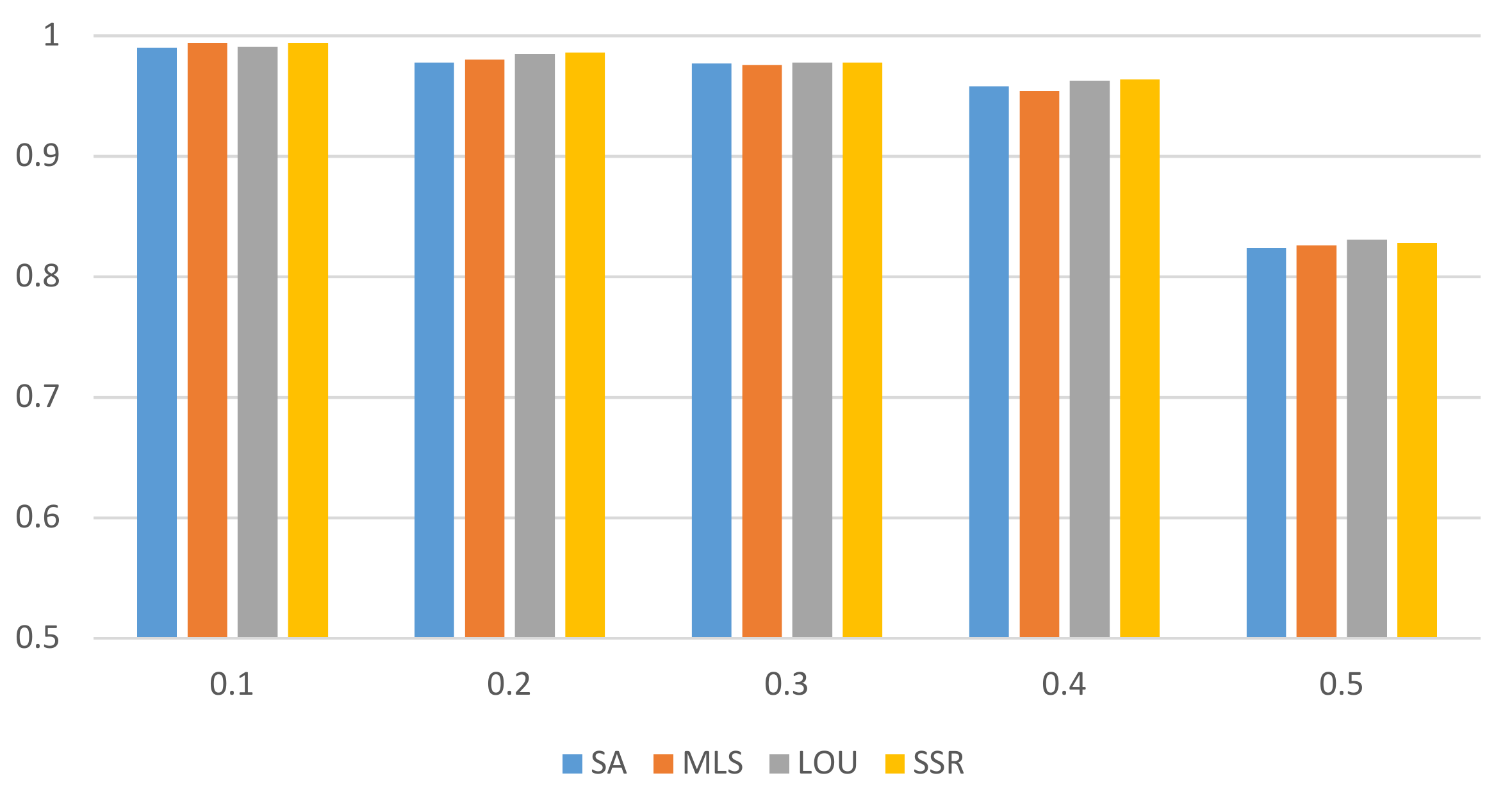}
    \caption{$n=1000, \bar{d}=10, \Delta=25$}\label{fig:nmi:a}    
  \end{subfigure}
  \begin{subfigure}[t]{2.5in}
    \centering
    \includegraphics[width=2.5in]{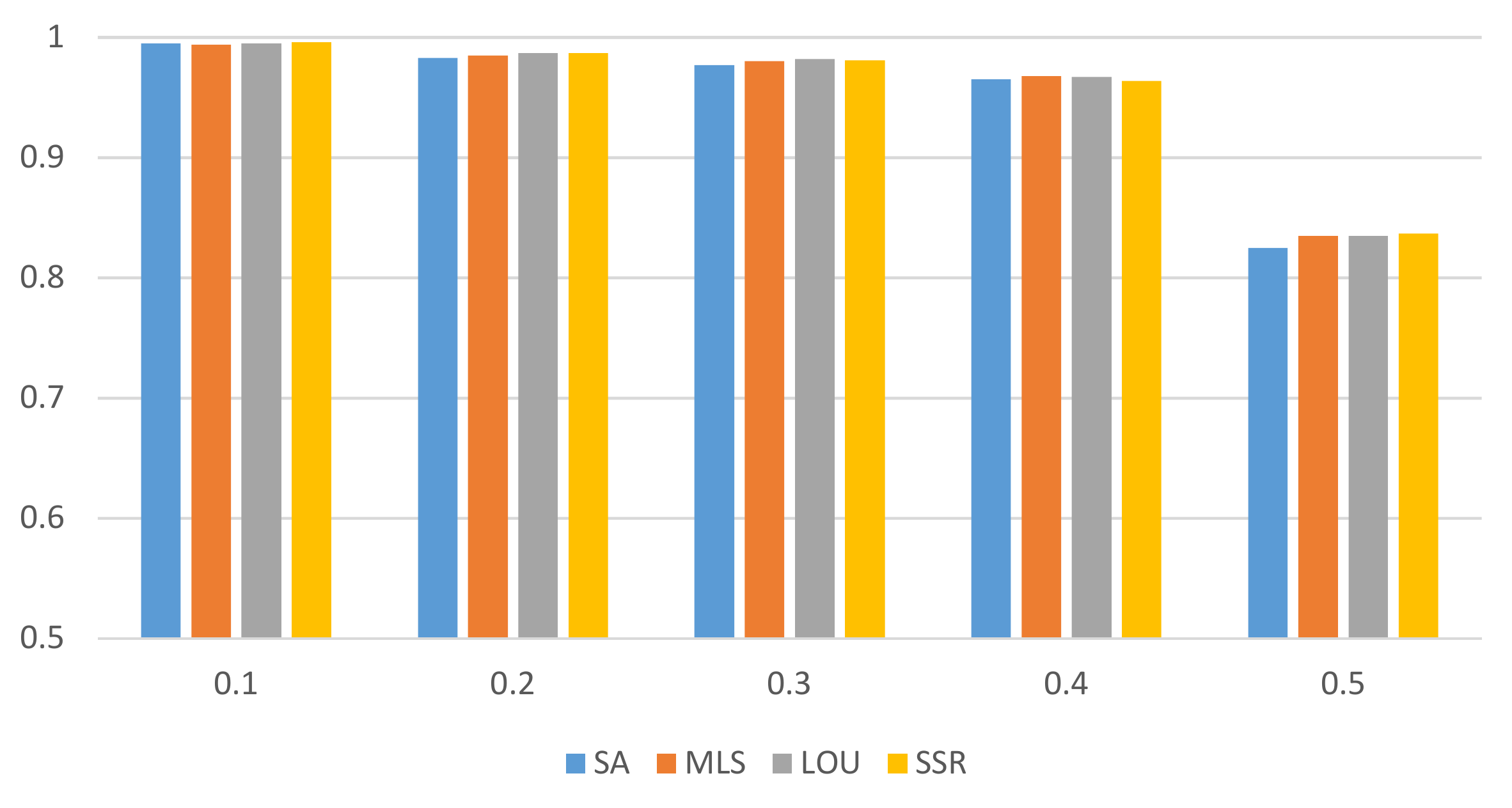}
    \caption{$n=1000, \bar{d}=20, \Delta=50$}\label{fig:nmi:b}
  \end{subfigure}\\
  \begin{subfigure}[t]{2.5in}
    \centering
    \includegraphics[width=2.5in]{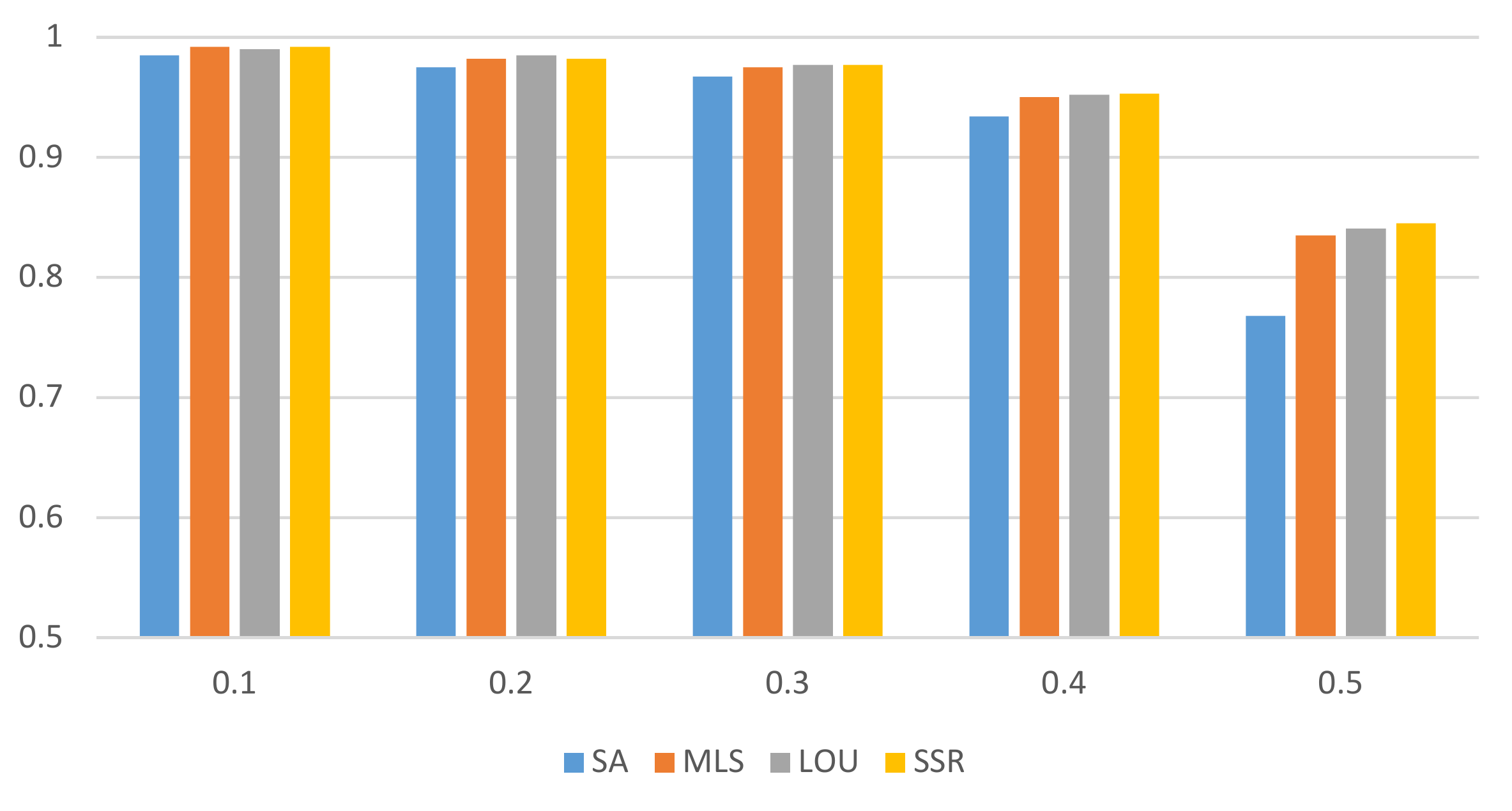}
    \caption{$n=5000, \bar{d}=10, \Delta=25$}\label{fig:nmi:c}    
  \end{subfigure}
  \begin{subfigure}[t]{2.5in}
    \centering
    \includegraphics[width=2.5in]{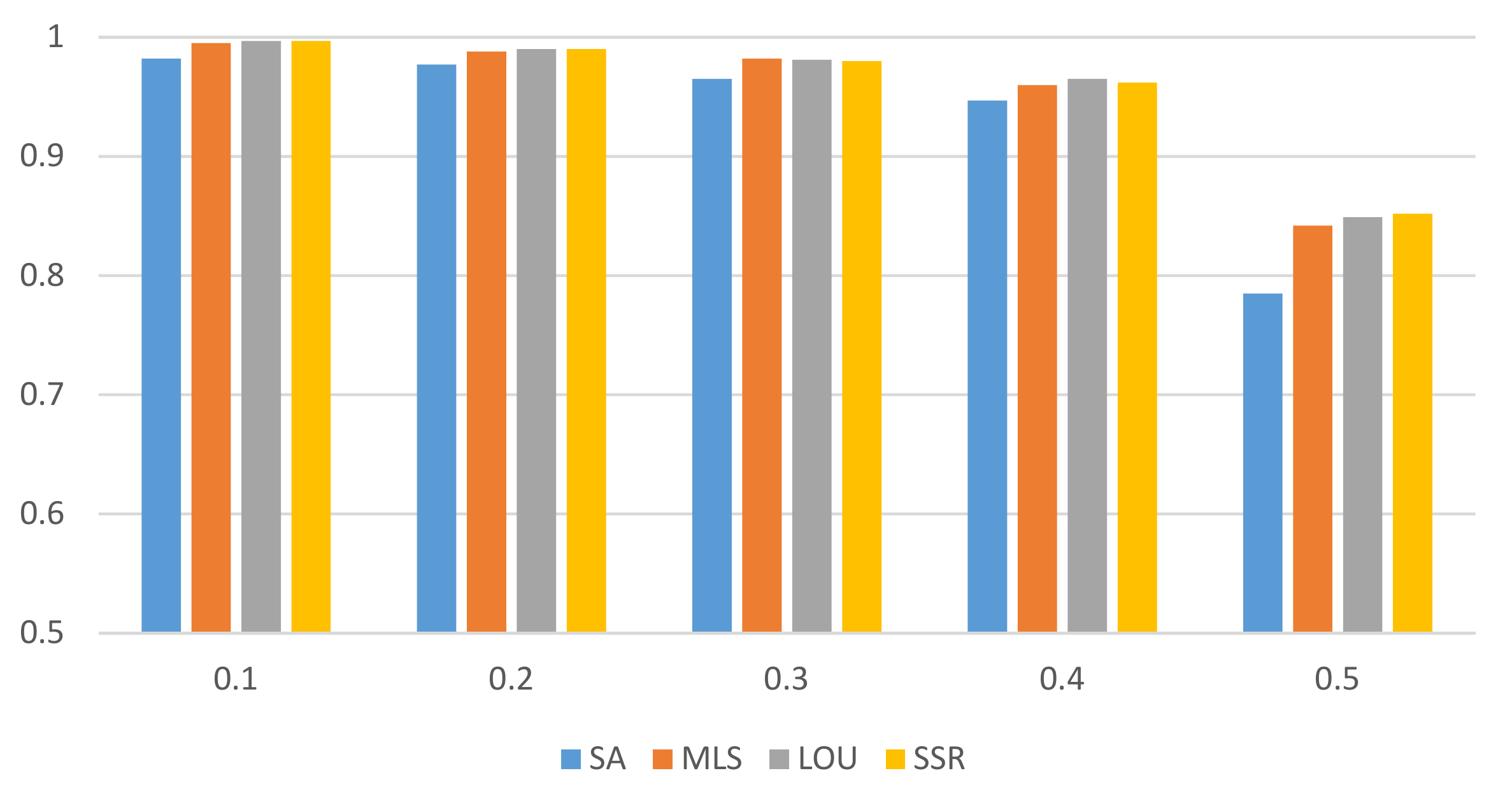}
    \caption{$n=5000, \bar{d}=20, \Delta=50$}\label{fig:nmi:d}
  \end{subfigure}
  \caption{Comparison of NMI values by different methods. Horizontal: mixing parameter values (0.1--0.5). Vertical: NMI values (0.5--1.0). (For computational concerns, SA used an annealing parameter value of $0.99$ on networks with $1,000$ vertices, and a value of $0.90$ on networks with $5,000$ vertices.)}
  \label{fig:nmi}
\end{figure}

Besides the modularity values, our second goal is on the method's sensitivity towards the change of network structures. We synthesized artificial networks under different structural settings. With known network structures, we are able to evaluate the performance of different methods in revealing the communities by comparing the results with the ground-truth.

In our experiments, twenty networks were generated with the LFR method \cite{LancichinettiFR:2008}. The networks have various number of vertices ($n=1000, 5000$), average degrees ($\bar{d}=10,20$), maximum degrees ($\Delta=25,50$) and mixing parameters ($0.1, 0.2, \cdots, 0.5$). A mixing parameter gives the ratio of inter-community edges over all edges. A parameter value of $0.5$ is the border beyond which the network community structures are not significant any more in the sense that the vertices have fewer intra-community connections than inter-community connections \cite{RadicchiCCLP:2004}.

The normalized mutual information measure, or $NMI$, is routinely applied to show the quality of community detection results when the true structure is known \cite{DanonDDA:2005}. Given the true partition $P_{A}$ and a candidate partition $P_{B}$, let $r_{a}$ be the number of communities in $P_{A}$ and $r_{b}$ be the number of communities in $P_{B}$. Let $n_{kk^{\prime}}$ be the number of vertices that appear in community $k$ of $P_A$ and also found in community $k^{\prime}$ of $P_B$. Denote $n_{k.}=\sum_{k^{\prime}}n_{kk^{\prime}}$ and $n_{.k^{\prime}}$ and $n_{.k^{\prime}}=\sum_{k}n_{kk^{\prime}}$. The $NMI$ measure quantifies the quality of the partition $P_{B}$ by:
\begin{equation}
NMI_{A,B} =\frac{-2\sum_{k=1}^{r_{a}}\sum_{k=1}^{r_{b}}n_{kk^{%
\prime }}\log \left( \frac{n_{kk^{\prime }}n}{n_{k.}n_{.k^{\prime }}}\right)
}{\sum_{k=1}^{r_{a}}n_{k.}\log \left( \frac{n_{k.}}{n}\right)
+\sum_{k^{\prime }=1}^{r_{b}}n_{.k^{\prime }}\log \left( \frac{n_{.k^{\prime
}}}{n}\right) }.
\end{equation}
The $NMI$ value lies in the range of $\left[0,1 \right]$. A larger the $NMI$ value indicates a higher quality of the candidate partition complying with the true partition. If the two partitions are identical, the $NMI$ value reaches $1$. If they completely independent, the $NMI$ value approaches $0$.

We compared the $NMI$ values for the SA, MLS, LOU and SSR methods on different networks. Fig. \ref{fig:nmi} shows the results. It can be seen that on these synthetic networks, all methods showed comparable results with high division quality on most of the networks. The SSR method had comparable sensitivity as the state-of-the-art approaches towards the change of network structures.

Evidently the mixing parameter plays a key role that affects the partition qualities. All four methods showed similar sensitivity patterns towards the change of this parameter. When its value is less than or equal to $0.4$, the $NMI$s are very near to $1$. When its value approaches $0.5$, however, there is an evident drop of the $NMI$ value as the community structure becomes too weak. The observed pattern is consistent with the trend revealed in a previous study \cite{LancichinettiFR:2008}.

\subsection{Parallel execution efficiency}
\label{sec:evaluation:speed}

\begin{figure}[!t]
  \centering
  \includegraphics[width=2.5in]{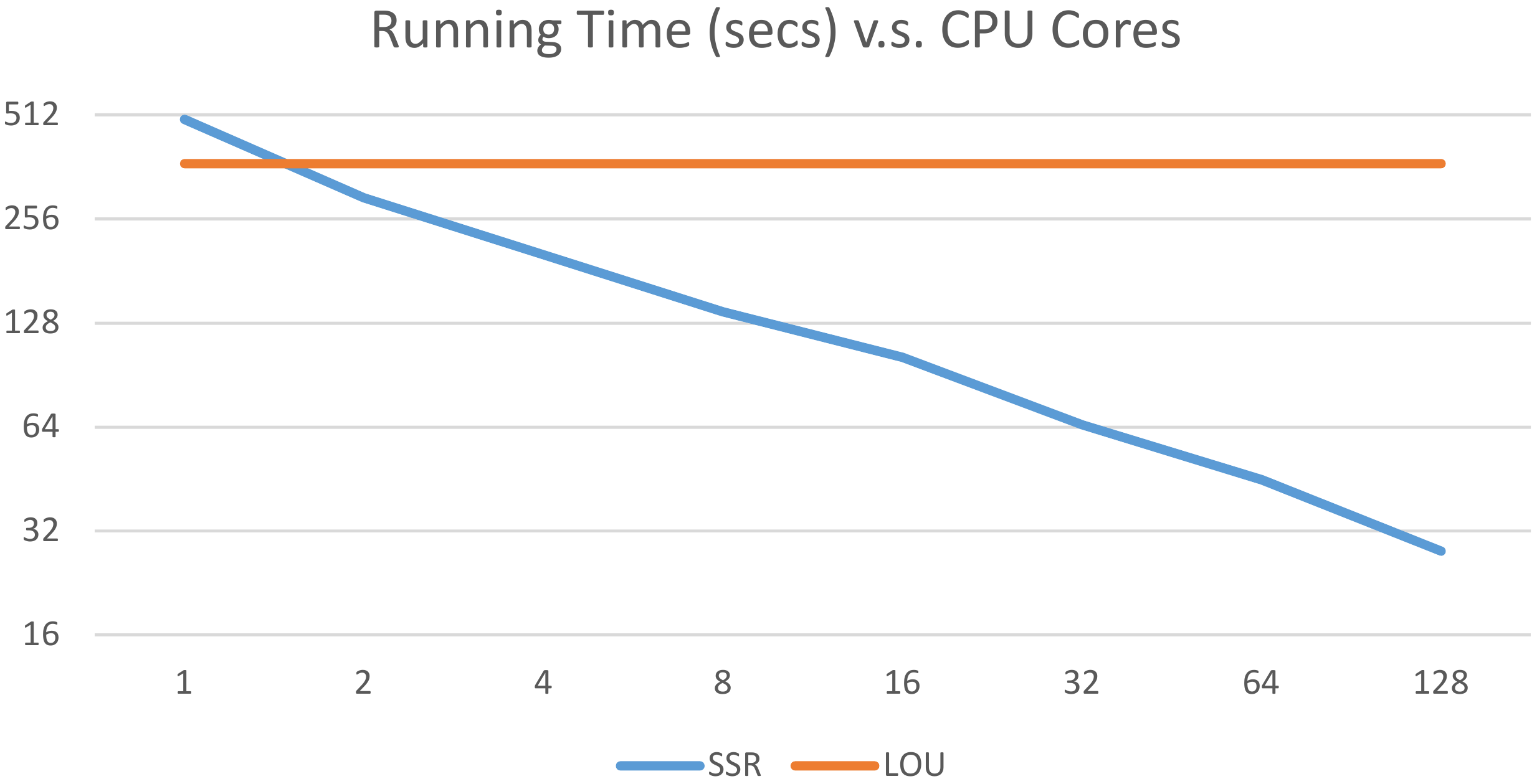}
  \caption{Horizontal: number of CPU cores; Vertical: Running time in seconds.}
  \label{fig:speed}
\end{figure}

Besides the qualities and the sensitivities, we also investigated the running time of the proposed method with different numbers of computing units, and compared the results. A large network, \textit{cit-Patents}, which has over three million vertices and sixteen million edges, was used in the evaluation.

The results are shown in Figure \ref{fig:speed}, where the horizontal axis gives the number of computing nodes (CPU cores), from a single node to $128$ nodes and the vertical axis shows the running time in seconds (log-scale). It can be seen that the execution time of the SSR method drops nearly linearly with the increase of computing nodes, from around $500$ seconds with one node to less than $30$ seconds with $128$ nodes, which verifies the high efficiency of the proposed SSR method when running in parallel computing platforms.

Comparatively, the LOU method spent around $350$ seconds, which is slightly faster than the SSR method (implemented in MATLAB) with one computing node. Unfortunately, the execution of the LOU method is not readily to be parallelized and benefits little from multiple computing nodes. An insightful inspection may find that the LOU method is an iterative method and each iteration has two phases: a local search phase and a network building phase. The two phases are highly dependent and couldn't be executed simultaneously. Besides, the major computation comes from the local search phase, within which a local exchange heuristic is repeated by moving one vertex from one community to another community, in a way similar to the Kernighan-Lin algorithm \cite{LinKernighan:1973}. The heuristic has strong dependence between consecutive exchanges and thus the operations in the first phase can't be executed simultaneously either.

\section{Conclusion}
\label{sec:conclusion}

With invaluable theoretical values and tremendous practical applications, the study of community detection and modularity maximization in complex networks has attracted much research attention recently. Unfortunately, the inherent NP-completeness nature of the problem poses non-trivial challenge and makes it difficult for most computational approaches to scale to large networks. To address the issue, we proposed a successive spectral relaxation based method to optimize the modularity measure. The key component of the proposed method is an algorithm that effectively finds the leading eigenvector of the modularity matrix while satisfying the required linear constraints. The method is simple and easy to implement. In benchmark evaluations it has reported high quality results comparable to the state-of-the-art approaches.

A highly notable feature of the proposed method is that it only involves basic matrix-vector multiplication, addition and normalization operations and runs in parallel computing platforms with very high efficiency. Empirically the proposed method shows a nearly linear speed-up with the increase of computing nodes. It divides a network with millions of vertices in tens of seconds time with $128$ CPU cores, a significant improvement over other approaches. Thereby the proposed method provides a highly promising and practical solution in detecting communities in very large networks.

\section{Acknowledgments}
\label{sec:acknowledgments}

This work is supported by Shenzhen Fundamental Research Fund under Grant No. KQTD2015033114415450 and Grant No. JCYJ20170306141038939.

\bibliographystyle{abbrv}
\bibliography{mybib}
\end{document}